\pdfoutput=1
\documentclass[a4paper,UKenglish]{lipics}
\usepackage{microtype}
\usepackage[T1]{fontenc}
\usepackage{amsthm}
\usepackage{setspace} 
\usepackage{xspace}
\usepackage{paralist}
\usepackage{enumerate}
\usepackage{comment}
\usepackage{url}
\usepackage{amssymb}
\usepackage{amsmath}
\usepackage{paralist}
\usepackage[active]{srcltx}
\usepackage{algorithm}
\usepackage{algpseudocode} 
\usepackage{bbm}
\usepackage{graphicx}
\graphicspath{{./Figs/}}
\usepackage{color} 
\usepackage{latexsym}

\newenvironment{proof sketch}[1]{\noindent {\emph{Proof sketch of #1:}}}{\hfill \qed}

\newtheorem{claim}{Claim}
\newtheorem{fact}{Fact}

\newtheorem{observation}{Observation}
\newtheorem{coro}{Corollary}

\newcommand{\etal}{\textit{et al.}}
\newcommand{\eqdf}{\triangleq}
\newcommand{\eps}{\varepsilon}

\newcommand{\cmin}{c_{\min}}

\newcommand{\bmax}{b_{\max}}
\newcommand{\dmax}{d_{\max}}
\newcommand{\Deltamax}{\Delta_{\max}}

\newcommand{\ppn}{\mbox{\rm pn}}

\newcommand{\alg}{\textsc{alg}\xspace}

\newcommand{\out}{\mathsf{out}}
\newcommand{\inn}{\mathsf{in}}

\newcommand{\Set}[1]{\left\{#1\right\}}

\newcommand{\namedref}[2]{\hyperref[#2]{#1~\ref*{#2}}}

\newcommand{\theoremref}[1]{\namedref{Theorem}{#1}}

\newcommand{\corollaryref}[1]{\namedref{Corollary}{#1}}

\newcommand{\Algref}[1]{\namedref{Algorithm}{#1}}

\newcommand{\equalityref}[1]{\hyperref[#1]{Equality~\eqref{#1}}}
\newcommand{\inequalityref}[1]{\hyperref[#1]{Inequality~\eqref{#1}}}
\newcommand{\factref}[1]{\namedref{Fact}{#1}}

\usepackage{color}
\usepackage{soul}

\renewcommand{\paragraph}[1]{\par\noindent\textbf{#1}}

\newcommand{\expec}[1]{\mathbf{E} \left[ {#1} \right]}
\newcommand{\PR}{{\sc pr}}
\newcommand{\pr}[1]{\mathbf{Pr} \left[ {#1} \right]}
\newcommand{\opt}{\mathsf{opt}}
\newcommand{\pcfp}{PCFP}
\newcommand{\mult}{\textsf{multiplicity}}
\newcommand{\bit}{\textit{bit}}

\title{An Approximation Algorithm for Path Computation and Function Placement
    in SDNs}
\titlerunning{Path Computation and Function Placement} 

\author[1]{Guy Even}
\author[2]{Matthias Rost}
\author[3]{Stefan Schmid}
\affil[1]{School of Electrical Engineering\\
 Tel Aviv University\\ Tel Aviv 6997801\\ Israel \\
  \texttt{guy@eng.tau.ac.il}}
\affil[2]{Technische Universit\"at Berlin \\ 	
10587 Berlin\\
Germany\\
\texttt{mrost@inet.tu-berlin.de}}
\affil[3]{Department of Computer Science\\
Aalborg University\\
DK-9220 Aalborg\\
Denmark\\
\texttt{schmiste@cs.aau.dk}}
\authorrunning{Even, Rost, and Schmid} 


\renewcommand{\copyrightline}{} 

\begin{document}

\maketitle
\begin{abstract}
  We consider the task of computing (combined) function mapping and
  routing for requests in Software-Defined Networks (SDNs).  Function
  mapping refers to the assignment of nodes in the substrate network
  to various processing stages that requests must undergo. Routing
  refers to the assignment of a path in the substrate network that
  begins in a source node of the request, traverses the nodes that are
  assigned functions for this request, and ends in a destination of
  the request. 

  The algorithm either rejects a request or completely serves a
  request, and its goal is to maximize the sum of the benefits of the
  served requests. The solution must abide edge and vertex capacities.

  We follow the framework suggested by Even~\etal\cite{EMPS15} for the
  specification of the processing requirements and routing of requests via
  processing-and-routing graphs (PR-graphs). In this framework, each
  request has a demand, a benefit, and PR-graph.

  Our main result is a randomized approximation algorithm for path computation and
  function placement with the following guarantee.  Let $m$ denote the number of
  links in the substrate network, $\eps$ denote a parameter such that $0< \eps <1$,
  and $\opt_f$ denote the maximum benefit that can be attained by a fractional
  solution (one in which requests may be partly served and flow may be split along
  multiple paths).  Let $\cmin$ denote the minimum edge capacity, and let $\dmax$
  denote the maximum demand. Let $\Deltamax$ denote an upper bound on the number of
  processing stages a request undergoes. If $\cmin/(\Deltamax\cdot\dmax)=\Omega((\log
  m)/\eps^2)$, then with probability at least
  $1-\frac{1}{m}-\textit{exp}(-\Omega(\eps^2\cdot \opt_f /(\bmax \cdot \dmax)))$, the
  algorithm computes a $(1-\eps)$-approximate solution.
\end{abstract}

\section{Introduction}
Software Defined Networks (SDNs) and Network Function Virtualization (NFV) have been
reinventing key issues in networking~\cite{Kreutz2015survey}. The key characteristics
of these developments are: (i)~separation between the data plane and the management
(or control) plane, (ii)~specification of the management of the network from a global
view, (iii)~introduction of network abstractions that provide a simple networking
model, and (iv)~programmability of network components.

In this paper we focus on an algorithmic problem that the network manager needs to
solve in an NFV/SDN setting. This problem is called \emph{path computation and
  function placing}. Path computation is simply the task of allocating paths to
requests. These paths are subject to the capacity constraints of the network links
and the forwarding capacity of the network nodes.  In modern networks, networking is
not limited to forwarding packets from sources to destinations. Requests can come in
the form of flows (i.e., streams of packets from a source node to a destination node
with a specified packet rate) that must undergo processing stages on their way to
their destination. Examples of processing steps include: compression, encryption,
firewall validation, deep packet inspection, etc. The crystal ball of NFV is the
introduction of abstractions that allow one to specify, per request, requirements
such as processing stages, valid locations for each processing stage, and allowable
sets of links along which packets can be sent between processing stages. An important
example for such goal is supporting security requirements that stipulate that
unencrypted packets do not traverse untrusted links or reach untrusted nodes.

From an algorithmic point of view, the problem of path computation and function
mapping combines two different optimization problems. Path computation alone (i.e.,
the case of pure packet forwarding without processing of packets) is an integral path
packing problem. Function mapping alone (i.e., the case in which packets only need to
be processed but not routed) is a load balancing problem. 

To give a feeling of the problem, consider a special case of requests for streams,
each of which needs to undergo the same sequence of $k$ processing stages
$w_1,w_2,\ldots, w_k$. This means that service of a request from $s_i$ to $t_i$ is
realized by a concatenation of $k+1$ paths: $s_i \overset{p_0}{\leadsto} v_1
\overset{p_1}{\leadsto} v_2 \overset{p_2}{\leadsto} \cdots
\overset{p_{k-1}}{\leadsto} v_k \overset{p_k}{\leadsto} t_i$, where processing stage
$w_i$ takes place in node $v_i$.  Note that the nodes $v_1,\ldots, v_k$ need no be
distinct and the concatenated path $p_0\circ p_1 \circ \cdots\circ p_k$ need not be
simple. A collection of allocations that serve a set of requests not only incurs a
forwarding load on the network elements, it also incurs a computational load on the
nodes. The computational load is created by the need to perform the processing stages
for the requests.

\paragraph{Previous works.}
Most papers on the topic resort to heuristics or non-polynomial algorithms. For
example, in~\cite{soule2014merlin} mixed-integer programming is employed.  The online
version is studied in~\cite{EMPS15} in which new standby/accept service model is
introduced. 

\paragraph{Contribution and Techniques.}
Under reasonable assumptions (i.e., logarithmic capacity-to-demand ratio and
sufficiently large optimal benefit), we present the first offline approximation
algorithm for the path computation and function placing problem.  Our starting point
is the model of SDN requests presented in~\cite{EMPS15}.  In this model, each request
is represented by a special graph, called a place-and-route graph (\PR-graph, in
short). The \PR-graph represents both the routing requirement and the processing
requirements that the packets of the stream must undergo. We also build on the
technique of graph products for representing valid realizations of
requests~\cite{EMPS15}. We propose a fractional relaxation of the problem. The
fractional relaxation consists of a set of fractional flows, each over a different
product graph. Each flow is fractional in the sense that it may serve only part of a
request and may split the flow among multiple paths. We emphasize that the fractional
flows do not constitute a multi-commodity flow because they are over different
graphs. Nevertheless, the fractional problem is a general packing
LP~\cite{raghavan1986randomized}.  We solve the fractional relaxation and apply
randomized rounding~\cite{raghavan1986randomized} to find an approximate solution.

Although randomized rounding is very well known and appears in many textbooks and
papers, the version for the general packing problem appears only in half a page in
the thesis of Raghavan~\cite[p. 41]{raghavan1986randomized}. A special case with unit
demands and unit benefits appears in~\cite{motwani1996randomized}.  Perhaps one of
the contributions of this paper is a full description of the analysis of randomized
rounding for the general packing problem.

\section{Modeling Requests in SDN}
In Even \etal~\cite{EMPS15}, a model for SDN requests, based on so called
place-and-route graphs (\PR-graphs) and product graphs is presented.  The model is
quite general, and allows each request to have multiple sources and destinations,
varying bandwidth demand based on processing stages, task specific capacities,
prohibited locations of processing, and prohibited links for routing between
processing stages, etc. We overview a simplified version of this model so that we can
define the problem of path computation and function placement.

\subsection{The Substrate Network}
The substrate network is a fixed network of servers and
communication links. The network is represented by a graph $N = (V,E)$, where $V$ is
the set of \emph{nodes} and $E$ is the set of \emph{edges}.
Nodes and edges have \emph{capacities}. The capacity of an edge $e$ is denoted by
$c(e)$, and the capacity of a node $v\in V$ is denoted by $c(v)$.  By scaling, we may
assume that $\min_{x\in V\cup E} c(x) = 1$.  We note that the network is static and undirected
(namely each edge represents a bidirectional communication link), but may contain
parallel edges.

\subsection{Requests and \PR-Graphs}
Each request is specified by a tuple $r_j = (G_j,d_j,b_j,U_j)$, where the components
are as follows:
\begin{enumerate}
\item $G_j=(X_j,Y_j)$ is a directed (acyclic) graph called the place-and-route graph
  (\PR-graph). There is a single source (respectively, sink) that corresponds to the
  source (resp.  destination) of the request. We denote the source and sink nodes in
  $G_j$ by $s_j$ and $t_j$, respectively. The other vertices correspond to services
  or processing stages of a request.  The edges of the \PR-graph are directed and
  indicate precedence relations between \PR-vertices.
\item The demand of $r_j$ is $d_j$ and benefit is $b_j$. By scaling, we may
  assume that $\min_j b_j =1$. 
\item $U_j:X_j\cup Y_j\to 2^V\cup 2^E$ where $U_j(x)$ is a set of ``allowed'' nodes
  in $N$ that can perform service $x$, and $U_j(y)$ is a set of ``allowed'' edges of
  $N$ that can implement the routing requirement that corresponds to $y$.
\end{enumerate}

\subsection{The Product Network}
For each request $r_j$, the product network $\ppn(N,r_j)$ is defined as follows.  The
node set of $\ppn(N,r_j)$, denoted $V_j$, is defined as $V_j\triangleq\cup_{y\in Y_j}
\left(U_j(y) \times \{y\}\right)$.  We refer to the subset $U_j(y)\times \{y\}$ as
the $y$-layer in the product graph.
The edge set of $\ppn(N,r_j)$, denoted $E_j$,
consists of two types of edges $E_j=E_{j,1}\cup E_{j,2}$ defined as follows.
\begin{enumerate}
\item\emph{Routing edges} connect vertices in the same layer.  
  \begin{align*}
    E_{j,1}&=\Set{\big((u,y),(v,y)\big) \mid y\in Y_j,\, (u,v)\in U_j(y)}~.
  \end{align*}

\item\emph{Processing edges} connect two copies of the same network vertex in
  different layers.
  \begin{align*}
    E_{j,2}&=\Set{\big((v,y),(v,y')\big) \mid y\neq y' \text{ are $2$ edges with a
        common endpoint $x$, and 
$v\in U_j(x)$}}.
  \end{align*}
\end{enumerate}

\subsection{Valid Realizations of SDN Requests}
Consider a path $\tilde{p}$ in the product graph $\ppn(N,r_j)$ that starts in the
$s_j$-layer and ends in the $t_j$-layer, where $s_j$ and $t_j$ are the source and
sink vertices of the \PR-graph $G_j$. Such a path $\tilde{p}$ represents the routing
of request $r_j$ from its origin to its destination and the processing stages that it
undergoes.  The processing edges along $\tilde{p}$ represent nodes in which
processing stages of $r_j$ take place. The routing edges within each layer represent
paths along which the request is delivered between processing stages.

\begin{definition}
  A path $\tilde{p}$ in the product network $\ppn(N,r_j)$ that starts in the (source)
  $s_j$-layer and ends in the (sink) $t_j$-layer is  a \emph{valid realization} of
  request $r_j$.
\end{definition}
We note that in~\cite{EMPS15} the projection of $\tilde{p}$ to the substrate network is
referred to as a valid realization.  The \emph{projection} of vertices of
$\ppn(N,r_j)$ to vertices in $N$ maps a vertex $(u,y)$ to $u$. By the definition of
the product graph, this projection maps paths in $\ppn(N,r_j)$ to paths in $N$.
Consider the path $p$ in $N$ resulting from the projection of a path $\tilde{p}$ in
the product graph. Note that $p$ may not be simply even if $\tilde{p}$ is simple. 

\subsection{The Path Computation and Function Placement Problem (\pcfp)}

\paragraph{Notation.}
Consider a path $\tilde{p}$ in the product graph $\ppn(N,r_j)$.  The
\emph{multiplicity} of an edge $e=(u,v)$ in the substrate network $N$ in $\tilde{p}$ is
the number of routing edges in $\tilde{p}$ that project to $e$, formally:
\begin{align*}
  \mult(e,\tilde{p}) &\triangleq | \{ y\in Y_j \mid ((u,y),(v,y)) \in \tilde{p}\}|
\end{align*}
Similarly, the multiplicity of a vertex $v\in V$ in $\tilde{p}$ is the number of
processing edges in $\tilde{p}$ that project to $v$, formally:
\begin{align*}
  \mult(v,\tilde{p}) &\triangleq | \{ y\in Y_j \mid \exists y': ((v,y),(v,y')) \in
  \tilde{p}\}|
\end{align*}

\paragraph{Capacity Constraints.}
Let $\tilde{P} = \{\tilde{p}_i\}_{i\in I}$ denote a set of valid realizations for a
subset $\{r_i\}_{i\in I}\subseteq R$ of requests. The set $\tilde{P}$ \emph{satisfies the
capacity constraints} if
\begin{align*}
  \sum_{i\in I} d_i\cdot\mult(e,\tilde{p}_i) &\leq c(e), ~~\text{for every edge $e\in
    E$}\\
  \sum_{i\in I} d_i\cdot\mult(v,\tilde{p}_i) &\leq c(v), ~~\text{for every vertex $v\in
    V$}
\end{align*}

\paragraph{Definition of the \pcfp-problem.}
The input in the \pcfp-problem consists of a substrate network $N=(V,E)$ and a set of
requests $\{r_i\}_{i\in I}$. The goal is to compute valid realizations
$\tilde{P}=\{\tilde{p}_i\}_{i\in I'}$ for a subset $I'\subseteq I$ such that:
\begin{inparaenum}[(1)]
\item $\tilde{P}$ satisfies the capacity constraints, and 
\item the benefit $\sum_{i\in I'} b_i$ is maximum.
\end{inparaenum}
We refer to the requests $r_i$ such that $i\in I'$ as the \emph{accepted} requests;
requests $r_i$ such that $i\in I\setminus I'$ are referred to as \emph{rejected}
requests.

\section{The Approximation Algorithm for \pcfp}
The approximation algorithm for the \pcfp-problem is described in this section. It is
a variation of Raghavan's randomized rounding algorithm for general packing
problems~\cite[Thm 4.7, p. 41]{raghavan1986randomized} (in which the approximation
ratio is $\frac{1}{e}-\sqrt{\frac{2\ln n}{\eps\cdot e \cdot \opt}}$ provided
that $\frac{\cmin}{\dmax} \geq \frac{\ln n}{\eps}$).

\subsection{Fractional Relaxation of the \pcfp-problem}
We now define the fractional relaxation of the \pcfp-problem. Instead of assigning a
valid realization $\tilde{p}_i$ per accepted request $r_i$, we assign a fractional
flow $\tilde{f}_i$ in the product graph $\ppn(N,r_i)$. The source of flow
$\tilde{f}_i$ is the source layer (i.e., a super source that is connected connected
to the all the nodes in the source layer). Similarly, the destination of
$\tilde{f}_i$ is the destination layer. The demand of $\tilde{f}_i$ is $d_i$ (hence
$|\tilde{f}_i| \leq d_i$). As in the integral case, the capacity constrains are
accumulated across all the requests.  Namely, let $f_i$ denote the projection of
$\tilde{f}_i$ to the substrate network. The edge capacity constraint for $e$ is
$\sum_{i} f_i(e) \leq c(e)$. A similar constraint is defined for vertex capacities.
The benefit of a fractional solution $F=\{f_i\}_i$ is $B(F)\triangleq\sum_i b_i \cdot
|f_i|$.

We emphasize that this fractional relaxation is not a multi-commodity flow. The
reason is that each $\tilde{f_i}$ is over a different product graph. However, the
fractional relaxation is a general packing LP.

\subsection{The Algorithm}
The algorithm uses a parameter $1>\epsilon>0$.
The algorithm proceeds as follows.
\begin{enumerate}
\item Divide all the capacities by $(1+\eps)$. Namely, $\tilde{c}(e)=c(e)/(1+\eps)$
  and $\tilde{c}(v)=c(v)/(1+\eps)$.
\item Compute an maximum benefit fractional \pcfp\ solution $\{\tilde{f}_i\}_i$.
\item Apply the randomized rounding procedure independently to each flow
  $\tilde{f}_i$ over the product network $\ppn(N,r_j)$. (See Appendix~\ref{sec:RR}
  for a description of the procedure). Let $p_i$ denote the path in $\ppn(N,r_i)$ (if
  any) that is assigned to request $r_i$ by the randomized rounding procedure. Let
  $\{f'_i\}_i$ denote the projection of $p_i$ to the substrate network. Note that
  each $f'_i$ is an unsplittable all-or-nothing flow. The projection of $p_i$ might
  not be a simple path in the substrate, hence the flow $f'_i(e)$ can be a multiple
  of the demand $d_i$.
\end{enumerate}

\subsection{Analysis of the algorithm}
  \begin{definition}
    The \emph{diameter} of a \PR-graph $G_j$ is the length of a longest path in $G_j$ from
    the source $s_j$ to the destination $t_j$. We denote the diameter of $G_j$ by $\Delta(G_j)$.
  \end{definition}
The diameter of $G_j$ is well defined because $G_j$ is acyclic for every request
$r_j$. In all applications we are sware of, the diameter $\Delta(G_j)$ is constant
(i.e., less than $5$).

\medskip
\paragraph{Notation.}
Let $\Deltamax\triangleq \max_{j\in I} \Delta(G_j)$ denote the maximum diameter of a
request. Let $\cmin$ denote the minimum edge capacity, and let $\dmax$ denote the
maximum demand.  Let $\opt_f$ denote a maximum benefit fractional \pcfp\ solution
(with respect to the original capacities $c(e)$ and $c(v)$).  Let $\alg$ denote the
solution computed by the algorithm. Let $B(S)$ denote the benefit of a solutions $S$.
Define $\beta(\eps)\eqdf (1+\eps)\ln (1+\eps) - \eps$.

\medskip
\noindent
Our goal is to prove the following theorem.\footnote{We believe there
  is a typo in the analogous theorem for integral MCFs with unit demands and unit benefits
  in~\cite[Thm 11.2, p. 452]{motwani1996randomized} and that a factor of $\eps^{-2}$
  is missing in their lower bound on the capacities.}
\begin{theorem}\label{thm:alg}
  Assume that $\frac{\cmin}{\Deltamax\cdot\dmax}\geq \frac{4.2+\eps}{\eps^2} \cdot
  (1+\eps)\cdot \ln |E|$ and $\eps\in (0,1)$.  Then,
  \begin{align}
\label{eq:caps}
\pr{\alg \text{ does not  satisfy the capacity constraints }} &\leq\frac{1}{|E|}\\
\label{eq:benefit}    
\pr{B(\alg) < \frac{1-\eps}{1+\eps}\cdot B(\opt_f)} & \leq e^{-\beta(-\eps) \cdot B(\opt_f)/(\bmax\cdot
  \dmax)}.
  \end{align}
\end{theorem}
We remark in asymptotic terms, the theorem states that if
$\frac{\cmin}{\Deltamax\cdot \dmax}=\Omega(\frac{\log |E|}{\eps^2})$, then $\alg$
satisfies the capacity constrains with probability $1-O(1/|E|)$ and attains a benefit
of $(1-O(\eps))\cdot B(\opt_f)$ with probability
$1-\textit{exp}(-\Omega(\eps^2)\opt_f/(\bmax\cdot \dmax))$.
\begin{proof}
  The proof is based on the fact that randomized rounding is applied to each flow
  $\tilde{f}_i$ independently. Thus the congestion of an edge in $\alg$ is the sum of
  independent random variables. The same holds for the $B(\alg)$. The proof proceeds
  by applying Chernoff bounds.

  Proof of Eq.~\ref{eq:caps}.  For the sake of simplicity we assume that there are no
  vertex capacities (i.e., $c(v)=\infty$).  The proof is based on the Chernoff bound
  in \theoremref{thm:Chernoff}. To apply the bound, fix a substrate edge $e\in E$.
Recall that $f'_i(e)$ is a flow path that is obtained by a projection of a path in
the product network $\ppn(N,r_i)$. Let
  \begin{align*}
    X_i&\triangleq \frac{f'_i(e)}{\Deltamax\cdot \dmax}\\
    \mu_i &\triangleq \frac{\tilde{c}(e)}{\Deltamax\cdot \dmax} \cdot \frac{\tilde{f}_i(e)}{\sum_{j\in
        I}\tilde{f}_j(e)}.
  \end{align*}
  The conditions of \theoremref{thm:Chernoff} are satisfied for the following
  reasons. Note that $0 \leq X_i \leq 1$ because $f'_i(e)\in
  \{0,d_i,\ldots,\Deltamax\cdot d_i\}$.  Also, by \namedref{Claim}{claim:expec},
  $\expec{X_i} = \tilde{f}_i(e)/(\Deltamax\cdot\dmax)$. Since $\sum_{j\in I}\tilde{f}_j(e) \leq
  \tilde{c}(e)$, it follows that $\expec{X_i}\leq \mu_i$. Finally, $\mu \triangleq \sum_{i\in I}
  \mu_i = \tilde{c}(e)/(\Deltamax\cdot \dmax)$.

  Let $\alg(e)$ denote the load incurred on the edge $e$ by $\alg$.  
Namely $\alg(e)\triangleq \sum_{i\in I} f'_i(e)$.
Note that $\alg(e)\geq (1+\eps)\cdot \tilde{c}(e)$ iff 
\begin{align*}
  \sum_{i\in I} X_i &\geq (1+\eps) \cdot \frac{\tilde{c}(e)}{\Deltamax\cdot\dmax} =
  (1+\eps)\cdot \mu.
\end{align*}
From
  \theoremref{thm:Chernoff} we conclude that:
\begin{align*}
  \pr{\alg(e) \geq (1+\eps)\cdot \tilde{c}(e)} &\leq e^{-\beta(\eps)\cdot \tilde{c}(e) /
    (\Deltamax\cdot \dmax)}
  \end{align*}
  By scaling of capacities, we have $c(e)=(1+\eps)\cdot \tilde{c}(e)$. By
  \factref{fact:b}, $\beta(\eps)\geq \frac{2\eps^2}{4.2+\eps}$. By the assumption
  $\frac{\tilde{c}(e)}{\Deltamax\dmax} \geq \frac{4.2+\eps}{\eps^2}\cdot \ln |E|$. We conclude
  that
\begin{align*}
  \pr{\alg(e) \geq c(e)} &\leq \frac{1}{|E|^2}.
  \end{align*}
Eq.~\ref{eq:caps} follows by applying a union bound over all the edges.

Proof of Eq.~\ref{eq:benefit}. The proof is based on the
Chernoff bound stated in \theoremref{thm:Chernoff2}. To apply the bound, let
  \begin{align*}
    X_i&\triangleq \frac{b_i\cdot |f'_i|}{\bmax \cdot \dmax}\\
    \mu_i &\triangleq \frac{b_i \cdot |\tilde{f}_i|}{\bmax \cdot \dmax}.
  \end{align*}
  The conditions of \theoremref{thm:Chernoff2} are satisfied for the following
  reasons.  Since $b_i \leq \bmax$ and $|f'_i|\leq \dmax$, it follows that $0\leq
  X_i\leq 1$.  Note that $\sum_i X_i = B(\alg)/(\bmax\cdot \dmax)$. By
  \corollaryref{coro:expec amount}, $\expec{X_i}=\mu_i$. Finally, by linearity,
  $\sum_i b_i\cdot |\tilde{f}_i| = \opt_f/(1+\eps)$ and $\sum_i \mu_i =
  \frac{B(\opt_f)}{(1+\eps)\bmax\cdot\dmax}$, and the theorem holds.
\end{proof}

\subsection{Unit Benefits}
We note that in the case of identical benefits (i.e., all the benefits equal one and hence
$\bmax=1$) one can strengthen the statement. 
If $B(\opt_f)> \cmin$, then the large capacities assumption implies that
$B(\opt_f)/(\dmax \cdot \bmax)\geq \cmin/\dmax \geq \eps^{-2}\cdot \ln |E|$.  This
implies that that $B(\alg)\geq (1-O(\eps))\cdot B(\opt_f)$ with probability at least
$1-1/poly(|E|)$.  By adding the probabilities of the two possible failures (i.e.,
violation of capacities and small benefit) and taking into account the prescaling of
capacities, we obtain that with probability at least $1-O(1/poly(|E|))$, randomized
rounding returns an all-or-nothing unsplittable multi-commodity flow whose benefit is
at least $1-O(\eps)$ times the optimal benefit.

\section{Discussion}
Theorem~\ref{thm:alg} provides an upper bounds of the probability that $\alg$ is not
feasible and that $B(\alg)$ is far from $B(\opt_f)$. These bounds imply that our
algorithm can be viewed as version of an asymptotic PTAS in the following sense.
Suppose that the parameters $\bmax$ and $\dmax$ are not a function of $|E|$. As the
benefit of the optimal solution $\opt_f$ increases, the probability that $B(\alg)\geq
(1-O(\eps))\cdot B(\opt_f)$ increases.  On the other hand, we need the
capacity-to-demand ratio to be logarithmic, namely, $\cmin \geq \Omega((\Deltamax\cdot\dmax\cdot \ln
|E|)/\eps^2)$. We believe that the capacity-to-demand ratio is indeed large in
realistic networks.  

\noindent 
\medskip
\paragraph{Acknowledgement.}
Research supported by the EU project UNIFY FP7-IP-619609.

\bibliography{randomized-rounding,sdn}

\begin{thebibliography}{1}

\bibitem{EMPS15}
Guy Even, Moti Medina, and Boaz Patt-Shamir.
\newblock Online path computation and function placement in {SDN}s.
\newblock ArXiv Technical Report 602.06169, 2015.

\bibitem{Kreutz2015survey}
D.~Kreutz, F.~M.~V. Ramos, P.~E. Verissimo, C.~E. Rothenberg, S.~Azodolmolky,
  and S.~Uhlig.
\newblock Software-defined networking: A comprehensive survey.
\newblock {\em Proceedings of the {IEEE}}, 103(1):14--76, 2015.

\bibitem{motwani1996randomized}
Rajeev Motwani, Joseph~Seffi Naor, and Prabhakar Raghavan.
\newblock Randomized approximation algorithms in combinatorial optimization.
\newblock In {\em Approximation algorithms for NP-hard problems}, pages
  447--481. PWS Publishing Co., 1996.

\bibitem{raghavan1986randomized}
Prabhakar Raghavan.
\newblock Randomized rounding and discrete ham-sandwich theorems: provably good
  algorithms for routing and packing problems.
\newblock In {\em Report UCB/CSD 87/312}. Computer Science Division, University
  of California Berkeley, 1986.

\bibitem{raghavan1987randomized}
Prabhakar Raghavan and Clark~D Tompson.
\newblock Randomized rounding: a technique for provably good algorithms and
  algorithmic proofs.
\newblock {\em Combinatorica}, 7(4):365--374, 1987.

\bibitem{soule2014merlin}
Robert Soul{\'e}, Shrutarshi Basu, Parisa~Jalili Marandi, Fernando Pedone,
  Robert Kleinberg, Emin~Gun Sirer, and Nate Foster.
\newblock Merlin: A language for provisioning network resources.
\newblock In {\em Proceedings of the 10th ACM International on Conference on
  emerging Networking Experiments and Technologies}, pages 213--226. ACM, 2014.

\bibitem{young1995randomized}
Neal~E Young.
\newblock Randomized rounding without solving the linear program.
\newblock In {\em SODA}, volume~95, pages 170--178, 1995.

\end{thebibliography}
\appendix
\section{Multi-Commodity Flows}\label{sec:MCF}

Consider a directed graph $G=(V,E)$. Assume that edges have non-negative capacities
$c(e)$.  For a vertex $u\in V$, let $\out(u)$ denote the outward neighbors, namely
the set $\{y \in V \mid (u,y)\in E\}$.  Similarly, $\inn(u) \triangleq \{ x\in V \mid
(x,u)\in E\}$.  Consider two vertices $s$ and $t$ in $V$ (called the \emph{source}
and \emph{destination} vertices, respectively).  A \emph{flow} from $s$ to $t$ is a
function $f:E \rightarrow \mathbb{R}^{\geq 0}$ that satisfies the following
conditions:
\begin{enumerate}[(i)]
\item Capacity constraints: for every edge $(u,v)\in E$, $0 \leq f(u,v)\leq c(u,v)$.
\item Flow conservation: for every vertex $u\in V\setminus\{s,t\}$ 
  \begin{align*}
\sum_{x\in \inn (u)} f(x,u) &= \sum_{y\in \out (u)} f(u,y).
  \end{align*}
\end{enumerate}
The \emph{amount} of flow delivered by the flow $f$ is defined by 
\begin{align*}
  |f| & \triangleq \sum_{y\in \out (s)} f(s,y) - \sum_{x\in \inn (s)} f(x,s).
\end{align*}

Consider a set ordered pairs of vertices $\{(s_i,t_i)\}_{i\in I}$.  An element $i\in
I$ is called a \emph{commodity} as it denotes a request to deliver flow from $s_i$ to
$t_i$.  Let $F\triangleq\{f_i\}_{i\in I}$ denote a set of flows, where each
flow $f_i$ is a flow from the source vertex $s_i$ to the destination vertex $t_i$.
We abuse notation, and let $F$ denote the sum of the flows, namely $F(e)\triangleq
\sum_{i\in I} f_i (e)$, for every edge $e$.  Such a sequence is a
\emph{multi-commodity flow} if, in addition it satisfies \emph{cumulative capacity
  constraints } defined by:
\begin{align*}
  \text{for every edge $(u,v)\in E$:} &~~~F(u,v) \leq c(u,v).
\end{align*}

Demands are used to limit the amount of flow per commodity. Formally, let
$\{d_i\}_{i\in I}$ denote a sequence of positive real numbers. We say that $d_i$ is the
\emph{demand} of flow $f_i$ if we impose the constraint that $|f_i|\leq d_i$. Namely,
one can deliver at most $d_i$ amount of flow for commodity $i$. 

The \emph{maximum benefit optimization problem} associated with multi-commodity flow
is formulated as follows. The input consists of a (directed) graph $G=(V,E)$, edge
capacities $c(e)$, a sequence source-destination pairs for commodities
$\{(s_i,t_i)\}_{i\in I}$. Each commodity has a nonnegative demand $d_i$ and benefit
$b_i$.  The goal is to find a multi-commodity flow that maximizes the objective
$\sum_{(u,v)\in E} b_i \cdot |f_i|$.  We often refer to this objective as the
\emph{benefit} of the multi-commodity flow.  When the demands are identical and the
benefits are identical, the maximum benefit problem reduces to a maximum
\emph{throughput} problem.

A multi-commodity flow is \emph{all-or-nothing} if $|f_i|\in \{0,d_i\}$, for every
commodity $i\in I$.  A multi-commodity flow is \emph{unsplittable} if the support of
each flow is a simple path. (The \emph{support} of a flow $f_i$ is the set of edges
$(u,v)$ such that $f_i(u,v)>0$.) We often emphasize the fact that a multi-commodity
flow is not all-or-nothing or not unsplittable by saying that it \emph{fractional}.

\section{Randomized Rounding Procedure}\label{sec:RR}
In this section we overview the randomized rounding procedure.
The presentation is based on~\cite{motwani1996randomized}.
Given an instance $F=\{f_i\}_{i\in I}$ of a fractional multi-commodity flow with
demands and benefits, we are interested in finding an all-or-nothing unsplittable
multi-commodity flow $F'=\{f'_i\}_{i\in I}$ such that the benefit of $F'$ is as close
to the benefit of $F$ as possible.  

\begin{observation}\label{obs:acyclic}
As flows along cycles are easy to eliminate, we
assume that the support of every flow $f_i\in F$ is acyclic.
\end{observation}

We employ a randomized procedure, called \emph{randomized rounding}, to obtain $F'$
from $F$. We emphasize that all the random variables used in the procedure are
independent.  The procedure is divided into two parts. First, we flip random
independent coins to decide which commodities are supplied. Next, we perform a
random walk along the support of the supplied commodities. Each such walk is a simple
path along which the supplied commodity is delivered.  We describe the two parts in
detail below.

\medskip
\paragraph{Deciding which commodities are supplied.}
For each commodity, we first
decide if $|f'_i|=d_i$ or $|f'_i|=0$.  This decision is made by tossing a biased
coin $\bit_i\in\{0,1\}$ such that
\begin{align*}
  \pr{\bit_i=1}&\triangleq\frac{|f_i|}{d_i}.
\end{align*}
If $\bit_i=1$, then we decide that $|f'_i|=d_i$ (i.e., commodity $i$ is fully supplied).
Otherwise, if $\bit_i=0$, then we decide that $|f'_i|=0$ (i.e., commodity $i$ is not
supplied at all). 

\medskip
\paragraph{Assigning paths to the supplied commodities.}
For each commodity $i$ that we decided to fully supply (i.e., $\bit_i=1$), we assign a simple
path $P_i$ from its source $s_i$ to its destination $t_i$ 
by following a random walk along the support of $f_i$. At each node, the random walk proceeds by rolling a dice. The probabilities of the sides of the dice are proportional to the flow amounts. A detailed description of the computation of the path $P_i$ is given in \Algref{alg:path}.

\begin{algorithm}
  \caption{Algorithm for assigning a path $P_i$ to flow $f_i$.}
\label{alg:path}
 \begin{algorithmic}[1]
\State $P_i\gets \{s_i\}$.
\State $u\gets s_i$
\While {$u\neq t_i$} \Comment{did not reach $t_i$ yet}
\State $v\gets \textit{choose-next-vertex}(u)$.
\State Append $v$ to $P_i$ 
\State $u\gets v$
\EndWhile
\State \textbf{return} $(P_i)$.

   \Procedure{$\textit{choose-next-vertex}$}{$u,f_i$} 
\Comment{Assume that $u$ is in the support of $f_i$}

   \State Define a dice $C(u,f_i)$ with $|\out (u)|$ sides.
   The side corresponding to an edge $(u,v)$ has probability
   $f_i{(u,v)} / (\sum_{(u,v') \in \out(u)} f_i(u,v'))$.

   \State Let $v$ denote the outcome of a random roll of the dice $C(u,f_i)$.

   \State \textbf{return $(v)$}
\EndProcedure
  \end{algorithmic}
\end{algorithm}

\medskip
\paragraph{Definition of $F'$.}
Each flow $f'_i\in F'$ is defined as follows. If $\bit_i=0$, then $f'_i$ is identically
zero. If $\bit_i=1$, then $f'_i$ is defined by
\begin{align*}
  f'_i(u,v) &\triangleq
  \begin{cases}
    d_i & \text{if $(u,v)\in P_i$,}\\
    0& \text{otherwise.}
  \end{cases}
\end{align*}
Hence, $F'$ is an all-or-nothing unsplittable flow, as required.
\section{Analysis of Randomized Rounding}\label{sec:RR analysis}
The presentation in this section is based on~\cite{motwani1996randomized}.
\subsection{Expected flow per edge}
\begin{claim} \label{claim:expec}
  For every commodity $i$ and every edge $(u,v)\in E$:
  \begin{align*}
    \pr{(u,v)\in P_i} &= \frac{f_i(u,v)}{d_i},\\
    \expec{f'_i(u,v)} &= f_i(u,v).
  \end{align*}
\end{claim}
\begin{proof}
  Since
\begin{align*}
  \expec{f'_i(u,v)}& = d_i \cdot \pr{(u,v)\in P_i},
\end{align*}
it suffices to prove the first part. 

An edge $(u,v)$ can belong to the path $P_i$
only if $f_i(u,v)>0$.  We now focus on edges in the support of $f_i$.  By
\namedref{Observation}{obs:acyclic}, the support is acyclic, hence we can sort the
support in topological ordering.  The claim is proved by induction on the position of
an edge in this topological ordering.

  The induction basis, for edges $(s_i,y)\in \out(s_i)$, is proved as follows. Since the
  support of $f_i$ is acyclic, it follows that $f_i(x,s_i)=0$ for every $(x,s_i)\in
  \inn(s_i)$. Hence $|f_i|=\sum_{y \in \out(s_i,f_i)}f_i(s_i,y)$.
Hence,
  \begin{align*}
    \pr{(s_i,y)\in P_i} &= \pr{\bit_i =1} \cdot \pr {\text{dice $C(s_i,f_i)$ selects $(s_i,y)$} \mid
      \bit_i=1}\\
    &= \frac{|f_i|}{d_i} \cdot \frac{f_i(s_i,y)}{\sum_{y \in
        \out(s_i,f_i)}f_i(s_i,y)}  \\
&=\frac{f_i(s_i,y)}{d_i},
  \end{align*}
and the induction basis follows.

The induction step, for an edge $(u,v)$ in the support of $f_i$ such that $u\neq
s_i$, is proved as follows.  Vertex $u$ is in $P_i$ if and only if $P_i$ contains an edge
whose head is $u$. We apply the induction hypothesis to these incoming
 edges, and use flow conservation to obtain 
  \begin{align*}
    \pr{u\in P_i} & = \pr {\bigcup_{x\in\inn(u)} (x,u)\in P_i}\\
&= \frac {1}{d_i} \cdot \sum_{x\in\inn(u)} f_i(x,u)\\
&=  \frac {1}{d_i} \cdot \left( \sum_{y\in\out(u)} f_i(u,y) \right).
 \end{align*}
Now,
  \begin{align*}
\pr{(u,v)\in P_i} & =\pr{u\in P_i} \cdot \pr{\text{dice $C(u,f_i)$ selects $(u,v)$}\mid u\in
      P_i}\\
    &= \frac {1}{d_i} \cdot \left( \sum_{y\in\out(u)} f_i(u,y)\right)
    \cdot \frac{f_i(u,v)}{\sum_{y\in\out(u)} f_i(u,y)}\\
    &=\frac{f_i(u,v)}{d_i},
  \end{align*}
and the claim follows.
\end{proof}

\medskip
\noindent
By linearity of expectation, we obtain the following corollary.
\begin{coro}\label{coro:expec amount}
  $\expec{|f'_i|} = |f_i|$.
\end{coro}

\section{Mathematical Background}
In this section we present material from Raghavan~\cite{raghavan1987randomized} and
Young~\cite{young1995randomized} about the Chernoff bounds used in the analysis of
randomized rounding.

\begin{fact}\label{fact:exp}
  $e^x \geq 1+x$ and $x\geq \ln (1+x)$ for $x>-1$.
\end{fact}

\begin{fact}\label{fact:newton}
  $(1+\alpha)^x \leq 1+\alpha\cdot x$, for $0\leq x \leq 1$ and $\alpha\geq -1$.
\end{fact}

\begin{fact}[Markov Inequality]\label{fact:Markov}
  For a non-negative random variable $X$ and $\alpha>0$, $\pr{X\geq \alpha} \leq \frac{\expec{X}}{\alpha}$.
\end{fact}

\begin{definition}
  The function $\beta:(-1,\infty) \rightarrow \mathbb{R}$ is defined by $\beta(\eps)\eqdf (1+\eps)\ln
  (1+\eps) - \eps$.
\end{definition}
\begin{fact}\label{fact:b} For $\eps$ such that $-1<\eps<1$ we have
  $\beta(-\eps) \geq \frac{\eps^2}{2} \geq \beta(\eps) \geq
  \frac{2\eps^2}{4.2+\eps}$. Hence, $\beta(-\eps)=\Omega(\eps^2)$ and
  $\beta(\eps)=\Theta(\eps^2)$.
\end{fact}

\begin{theorem}[Chernoff Bound]\label{thm:Chernoff}
Let $\{X_i\}_i$ denote a sequence of independent random variables attaining values in $[0,1]$.
Assume that $\expec{X_i}\leq \mu_i$. Let $X\eqdf \sum_i X_i$ and $\mu\eqdf\sum_i \mu_i$.
Then, for $\eps >0$, 
\begin{align*}
  \pr{X\geq (1+\eps)\cdot \mu} &\leq e^{-\beta(\eps)\cdot \mu}.
\end{align*}
\end{theorem}
\begin{proof}
Let $A$ denote the event that $X\geq (1+\eps)\cdot \mu$.
Let $f(x)\eqdf (1+\eps)^x$.
Let $B$ denote the event that 
  \begin{align*}
\frac{f(X)}{f((1+\eps)\cdot \mu)}&\geq 1.
  \end{align*}
Because $f(x)>0$ and $f(x)$ is monotone increasing, it follows that $\pr{A}=\pr{B}$.
By Markov's Inequality, 
\begin{align*}
  \pr{B} &\leq \frac{\expec{f(X)}}{f((1+\eps)\cdot \mu)}.
\end{align*}
Since $X=\sum_i X_i$ is the sum of independent random variables,
\begin{align*}
  \expec{f(X)} &=\prod_i \expec{(1+\eps)^{X_i}}&\\
&\leq \prod_i \expec{1+\eps\cdot X_i} & (by~\factref{fact:newton})\\
&\leq \prod_i (1+\eps\cdot \mu_i) &\\
&\leq \prod_i e^{\eps\cdot \mu_i} & (by~\factref{fact:exp})\\
&= e^{\eps\cdot \mu} &
\end{align*}
We conclude that 
\begin{align*}
  \pr{A}&\leq \frac{e^{\eps\cdot \mu} }{f((1+\eps)\cdot \mu)}\\
&=e^{-\beta(\eps)\cdot \mu},
\end{align*}
and the theorem follows.
\end{proof}

We prove an analogue theorem for bounding the probability of the event that $X$ is much smaller than $\mu$.
\begin{theorem}[Chernoff Bound] \label{thm:Chernoff2}
Under the same premises as in~\theoremref{thm:Chernoff}
except that $\expec{X_i}\geq \mu_i$, it holds that, for $1> \eps \geq 0$, 
\begin{align*}
  \pr{X\leq (1-\eps)\cdot \mu} &\leq e^{-\beta(-\eps)\cdot \mu}.
\end{align*}
\end{theorem}
\begin{proof}
We repeat the proof of~\theoremref{thm:Chernoff} with the required modifications.
Let $A$ denote the event that $X\leq (1-\eps)\cdot \mu$.
Let $g(x)\eqdf (1-\eps)^x$.
Let $B$ denote the event that 
  \begin{align*}
\frac{g(X)}{g((1-\eps)\cdot \mu)}&\geq 1.
  \end{align*}
Because $g(x)>0$ and $g(x)$ is monotone decreasing, it follows that $\pr{A}=\pr{B}$.
By Markov's Inequality, 
\begin{align*}
  \pr{B} &\leq \frac{\expec{g(X)}}{g((1-\eps)\cdot \mu)}.
\end{align*}
Since $X=\sum_i X_i$ is the sum of independent random variables,
\begin{align*}
  \expec{g(X)} &=\prod_i \expec{(1-\eps)^{X_i}}&\\
&\leq \prod_i \expec{1-\eps\cdot X_i} & (by~\factref{fact:newton})\\
&\leq \prod_i (1-\eps\cdot \mu_i) &\\
&\leq \prod_i e^{-\eps\cdot \mu_i} & (by~\factref{fact:exp})\\
&= e^{-\eps\cdot \mu} &
\end{align*}
We conclude that 
\begin{align*}
  \pr{A}&\leq \frac{e^{-\eps\cdot \mu} }{g((1-\eps)\cdot \mu)}\\
&=e^{-\beta(-\eps)\cdot \mu},
\end{align*}
and the theorem follows.
\end{proof}

\end{document}